\documentclass{article}

\usepackage{microtype}
\usepackage{graphicx}
\usepackage{subfigure}
\usepackage{booktabs} % for professional tablese
\usepackage[most]{tcolorbox}

\usepackage{hyperref}

\usepackage[accepted]{icml2025}

\usepackage{amsmath}
\usepackage{amssymb}
\usepackage{mathtools}
\usepackage{amsthm}
\usepackage{graphicx}
\usepackage{booktabs}
\usepackage{fontawesome5}
\usepackage[capitalize,noabbrev]{cleveref}

\theoremstyle{plain}

\theoremstyle{definition}

\theoremstyle{remark}

\usepackage[textsize=tiny]{todonotes}

\icmltitlerunning{Personalized Artwork Recommendation}

\begin{document}

\twocolumn[
\icmltitle{Netflix Artwork Personalization via LLM Post-training}




\begin{icmlauthorlist}
\icmlauthor{Hyunji Nam}{Stanford}
\icmlauthor{Sejoon Oh}{Netflix}
\icmlauthor{Emma Kong}{Netflix}
\icmlauthor{Yesu Feng}{Netflix}
\icmlauthor{Moumita Bhattacharya}{Netflix}
\end{icmlauthorlist}

\icmlaffiliation{Stanford}{Stanford University}
\icmlaffiliation{Netflix}{Netflix}

\icmlcorrespondingauthor{Hyunji Nam}{hjnam@stanford.edu}

\icmlkeywords{Machine Learning, ICML}

\vskip 0.3in
]



\printAffiliationsAndNotice{}  % leave blank if no need to mention equal contribution

\begin{abstract}
Large language models (LLMs) have demonstrated success in various applications of user recommendation and personalization across e-commerce and entertainment~\citep{lyu-etal-2024-llm, lin2024recommendersystemsbenefitlarge, yang2023palrpersonalizationawarellms, Dai_2023, 10.1145/3631700.3665185, gao2023chatrecinteractiveexplainablellmsaugmented}. On many entertainment platforms such as Netflix, users typically interact with a wide range of titles, each represented by an artwork. Since users have diverse preferences, an artwork that appeals to one type of user may not resonate with another with different preferences. Given this user heterogeneity, our work explores the novel problem of personalized artwork recommendations according to diverse user preferences. Similar to the multi-dimensional nature of users' tastes, titles contain different themes and tones that may appeal to different viewers. For example, the same title might feature both heartfelt family drama and intense action scenes. Users who prefer romantic content may like the artwork emphasizing emotional warmth between the characters, while those who prefer action thrillers may find high-intensity action scenes more intriguing. Rather than a one-size-fits-all approach, we conduct post-training of pre-trained LLMs to make personalized artwork recommendations, selecting the most preferred visual representation of a title for each user and thereby improving user satisfaction and engagement. Our experimental results with Llama 3.1 8B~\citep{weerawardhena2025llama31foundationaisecurityllm8binstructtechnicalreport} models (trained on a dataset of 110K data points and evaluated on 5K held-out user-title pairs) show that the post-trained LLMs achieve 3-5\% improvements over the Netflix production model, suggesting a promising direction for granular personalized recommendations using LLMs. 
\end{abstract}

\section{Introduction} Large language models (LLMs) have shown success in various applications of user recommendation and personalization across e-commerce and entertainment~\citep{lyu-etal-2024-llm, lin2024recommendersystemsbenefitlarge, yang2023palrpersonalizationawarellms, Dai_2023, 10.1145/3631700.3665185, gao2023chatrecinteractiveexplainablellmsaugmented}. Users are typically represented through natural language descriptions of their backgrounds, demographics, or relevant interaction histories, and LLMs generate personalized predictions of user behavior or optimal actions that align with user objectives, such as improving retention or satisfaction with the service~\citep{soni2023large, chen2024large}. Following this line of research, we explore the novel task of personalized artwork recommendations tailored to individual users' tastes and preferences. On entertainment platforms such as Netflix, users interact with a wide range of titles, each represented by an image (artwork). These artworks are often the first and important cues that influence users' decision to watch or skip a show and therefore have been an important factor of content personalization~\citep{netflix-blog}. As users have diverse preferences, an artwork that appeals to one type of user may not resonate with another. Therefore, representing every title by a single artwork may fail to capture the varied interest of different users, especially across age demographics, cultures, and languages. 

We extend the well-studied topic of text-based personalized recommendations with LLMs to a novel sub-problem: predicting personalized user preferences among multiple visual representations of a title. We leverage the advanced capabilities of LLMs to process long (input) contexts and consider long user interaction histories (e.g., movies they have watched and liked). While prior works have focused on personalized title recommendation~\citep{yang2023palrpersonalizationawarellms, wu2025rlpfreinforcementlearningprediction} (for example, using the 1-million MovieLens dataset~\citep{harper2015movielens}), we focus on a related sub-task of predicting which artwork, from a set of candidate images, would appeal most to specific users once the personalized title is selected. The number of artwork options varies by titles, ranging from as few as four to more than forty, with different degrees of variation. Our experiments with Llama 3.1-8B~\citep{weerawardhena2025llama31foundationaisecurityllm8binstructtechnicalreport} show that LLMs post-trained with supervised fine-tuning (SFT) with reasoning distillation from a larger model (e.g., Qwen 3-32B~\citep{qwq32b}) and Direct Preference Optimization (DPO) achieve performance improvements of 5\% and 3\%, respectively, on a held-out user-title set compared to the Netflix production model. These empirical results suggest a promising path for using LLMs in fine-grained, personalized content recommendations (e.g., artwork, synopsis, trailer) to engage with diverse user interests.

\begin{figure}[h]
  \centering
  \includegraphics[width=\linewidth]{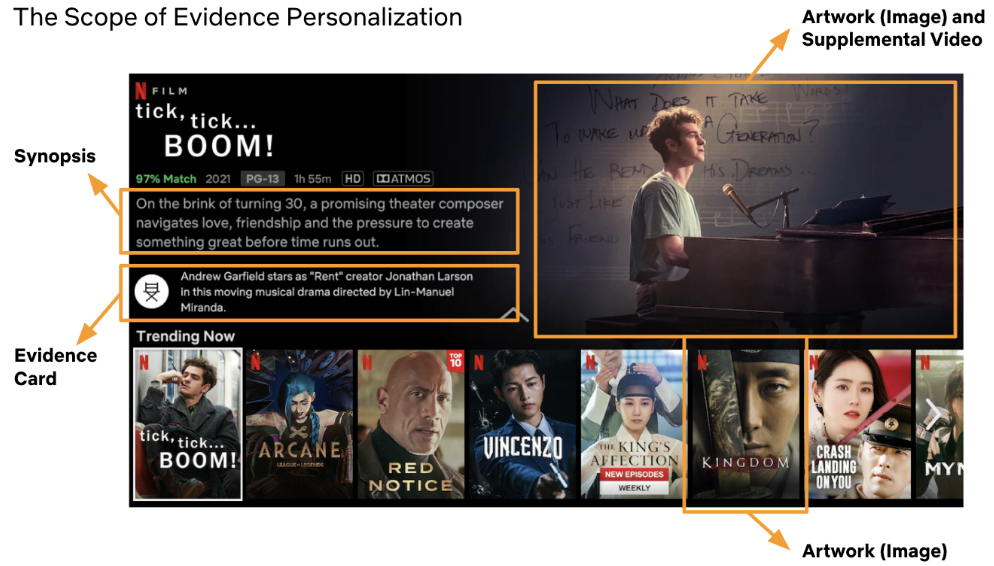}
  \caption{Different components of user experience (e.g., artworks, synopsis, trailers) can be personalized in addition to the title.}\label{fig:motivation}
\end{figure} 

\section{Related work}
\paragraph{LLMs for personalized recommendations} Numerous works have explored using LLMs for personalizing content and product recommendations to overcome the limitations of conventional recommendation models (e.g., lack of world knowledge, limited text understanding and reasoning capabilities), and have been successful~\citep{lyu-etal-2024-llm, lin2024recommendersystemsbenefitlarge}. ~\citet{yang2023palrpersonalizationawarellms} leverages user-item interaction history expressed in natural language to personalize movie and product recommendations with LLMs. 
~\citet{Dai_2023} evaluates ChatGPT's capabilities for point-wise (i.e., predicting a user's rating for a single product), pair-wise (i.e., predicting a user's preference between an item pair), and list-wise (i.e., ranking a list of candidate items based on the user's previous examples) recommendations across movies, books, music, and news datasets without any post-training. ~\citet{gao2023chatrecinteractiveexplainablellmsaugmented} explores the possibility of combining interactive LLM agents with traditional recommendation systems for cold-start and cross-domain generalizations, as well as using LLMs to provide explanations for a model's personalized selections. ~\citet{10.1145/3631700.3665185} specifically focuses on LLMs' abilities to generate explanations to improve transparency and engender user's trust in recommendation systems that may otherwise appear black-box. Unlike the prior works' focus on prompting and in-context learning from users' prior interaction examples, ~\citet{wu2025rlpfreinforcementlearningprediction} proposes training a summarizer that can produce user-specific summaries with reinforcement learning from prediction feedback (RLPF) and using the generated summaries as distilled user representations to personalize future product recommendations. Our work adds to this long line of research with a unique contribution: While prior works often include evaluations of movie and book recommendations, to our knowledge, they do not test the capabilities of LLMs to \textbf{personalize artwork selection for a given title}, which is unique to our work. 

\paragraph{LLM post-training} Post-training methods have been successfully applied to improve and refine the existing capabilities of LLMs trained on large-scale data~\citep{kumar2025llmposttrainingdeepdive, fernando2025understandingforgettingllmsupervised}. Popular strategies for post-training include supervised fine-tuning (SFT) and preference learning through reinforcement learning from human feedback (RLHF)~\citep{ouyang2022traininglanguagemodelsfollow, bai2022traininghelpfulharmlessassistant} or direct preference optimization (DPO)~\citep{rafailov2024directpreferenceoptimizationlanguage}. Knowledge distillation has also been widely explored as a technique for training smaller models (e.g., open-source Llama models) to adopt the capabilities of larger models, including proprietary models like GPT-4~\citep{gu2025minillmknowledgedistillationlarge, xu2024surveyknowledgedistillationlarge}, where the training data is typically generated by the larger model and provided to the smaller model for SFT. Other works have explored training language models with reasoning traces~\citep{zelikman2022starbootstrappingreasoningreasoning, huang-chang-2023-towards}, such as having models explain their predictions for commonsense question-answering tasks~\citep{rajani-etal-2019-explain} and generate step-by-step solutions to math problems~\citep{hendrycks2021measuringmathematicalproblemsolving}.

\paragraph{Automated image captioning} There are various approaches to representing images through text captions~\citep{li2023decapdecodingcliplatents, mokady2021clipcapclipprefiximage}, with more recent work based on multimodal large language models~\citep{meta-llama3.2} that can perform visual question answering~\citep{guo2023images}, reasoning~\citep{alayrac2022flamingovisuallanguagemodel}, and image captioning even in zero-shot settings and further improve performance with instruction tuning~\citep{li2025ottermultimodalmodelincontext, li2023mimicitmultimodalincontextinstruction, chen2023largelanguagemodelsvisual}.

\section{Problem setup} 
\subsection{Data composition} We have a dataset $\mathcal D$ consisting of users $U$, titles $X$, and multiple artwork options $A_{1:m}(X)$ for each title. Users are represented by their interaction histories which may include the user's viewing history and watch times, and the number of artwork options $m$  varies across titles. For each user-film pair, we also have ground-truth optimal artwork $A^*$ obtained from the user's previous engagement with the selected title.  All artworks other than $A^*$ are treated as negative examples that the user does not prefer. In our personalized recommendation setting, the optimal artwork depends on both the title and the user, $A^*(U, X)$.

For a test dataset $\mathcal D'$, we have a new set of user and title pairs, and our goal is to predict $A^*(U, X)$ for new $(U, X) \in \mathcal D'$ . While the test set could, in principle, include both held-out users and held-out titles, which means both components of personalization are unseen during training, we focus on cases where each user-title tuple is unique. In other words, the same user or the same title may appear in the training set, but never as the exact combination. 
% This step still ensures that the test dataset differs from the training set, since personalization depends on the specific interaction between a user and a film, making each $(U, X)$ a unique input. We leave ablations on held-out users, held-out movies, and their joint effects to future work.

\subsection{Evaluation metric}
We consider two metrics: \textbf{accuracy} and \textbf{inverse propensity score (IPS)}. While accuracy is more intuitive to interpret and broadly applicable, IPS is better suited for our setting, where the number of artwork options $m$ varies by titles. Accuracy is defined as:

\begin{equation}
 \sum_{i \in \mathcal D'} \frac{1}{|\mathcal D'|} I\{\hat A^*(u_i, x_i) = a^*_i\},
\end{equation} where $\hat A^*$ is the model's prediction for the optimal personalized artwork for a given tuple $(u_i, x_i)$. One limitation with accuracy is that it does not distinguish between cases with 40 artwork options versus 2 options. Intuitively, selecting the optimal artwork from a set of two has a higher chance of being correct by random selection compared to selecting the optimal artwork from a larger set. Therefore, a more nuanced evaluation metric would account for the number of artworks available for each title and appropriately reflect the difficulty of a correct selection by chance. 

This leads to IPS, which is defined as:
\begin{equation}
\sum_{i \in \mathcal D'} \frac{1}{|\mathcal D'|} \frac{I\{\hat A^*(u_i, x_i) = a^*_i\}}{\pi(a_i^*)}.
\end{equation} $\pi(a_i^*)$ denotes the probability of the ground truth artwork $a_i^*$ being shown to the user. By setting $\pi$ to be uniform over the set of artworks, we now have a metric that accounts for the difficulty of selecting the correct artwork from sets of different sizes. For example, a correct prediction from a set of 40 artwork candidates is weighted twenty times higher than a correct prediction from a set of two. Due to the varying number of artwork options across different films in the catalog, IPS is the primary evaluation metric used by the production system.

\section{Method}
\subsection{Problem formulation} For the user representation $X$, we use the user's most recent $K$ timestamped interactions, which may include information like the user's recently engaged titles and their genres. Each artwork is represented by a caption (of $\sim$ 200 tokens) generated by a fine-tuned Llama-3.2-11B visual language model~\citep{meta-llama3.2}. We formulate the task as a prediction problem: given the user's past interactions, a new title, and a list of artworks represented by captions, predict which artwork would most appeal to the user. Representing the images as captions enables all multimodal inputs to be expressed in text, which is easily supported by existing off-the-shelf LLMs for post-training, and the compactness of the text captions also allows our method to scale to a larger number of candidate artworks (e.g., 40+ options for certain titles).

% Therefore, both the user information $X$ and the artworks $A_{1:m}$ are expressed in text and can be used as inputs to the recommendation model.

% There are different possibilities of selecting the optimal personalized artwork for each user from a given set. One approach is to predict a preference score for each artwork independently and select the one with the highest score. Another approach is to provide all available artwork options to the model and have it predict the most preferred artwork. We focus on the latter approach because we do not score user preferences for non-preferred artworks and therefore, we have a severe class imbalance between positive and negative examples. However, if more fine-grained preference scores were available for each user, the first approach would be a reasonable alternative to consider. 

We conduct post-training of Llama 3.1 8B models~\citep{weerawardhena2025llama31foundationaisecurityllm8binstructtechnicalreport}. The prompt describes the user's interaction history and lists a set of artwork options for a new title, and the model's objective is to predict which artwork the user is most likely to prefer based on their taste and interest. 

\begin{tcolorbox}[colback=blue!5!white, colframe=blue!75!black, title=Personalized recommendation model instruction]
\textbf{Sample prompt template:} ``You are an expert in movies and shows. I want you to predict which of the available artworks the user would like the most based on their past watch history. [User's past interaction data] The user's new title is: [new title]. Here are the artwork options: \texttt{<option>} caption describing $A_1$ \texttt{</option>},  \texttt{<option>} caption describing $A_2$ \texttt{</option>}, ... ,\texttt{<option>} caption describing $A_m$ \texttt{</option>}. Output the best artwork in text.

\textbf{Model output:} ``Prediction: \texttt{<option>} caption describing $\hat A^*$ \texttt{</option>}.''
\end{tcolorbox}

We introduce two new tokens, \texttt{<option>} and \texttt{</option>}, to delimit different artwork options in the prompt. During inference, we guide the model's generation with the prefix ``Prediction: \texttt{<option>}.'' Additionally to extract the predicted caption, we use n-gram matching to the model's generation to select the artwork option with the highest match as the model's choice.

\subsection{Post-training method}
\begin{itemize}
\item \textbf{Supervised fine-tuning (SFT).} We use $(u_i, x_i, a^*_i)$ tuples from the training dataset (n=110K) to fine-tune Llama 3.1 8B-Instruct~\citep{weerawardhena2025llama31foundationaisecurityllm8binstructtechnicalreport} using the prompt and expected model output shown above. The SFT loss is defined as: \begin{equation}
\mathcal L_{\text{SFT}}(\pi_\theta) = - \mathbb E_{i \sim \mathcal D} \Bigg[ \log \pi_\theta(a_i^*|x_i, u_i) \Bigg]
\end{equation}

\item \textbf{Prediction with reasoning.} We leverage the recently advanced reasoning capabilities of LLMs to augment personalized selection. Specifically, we augment the training data leveraging the powerful reasoning model, Qwen/QwQ-32B~\citep{qwq32b}, to generate explanations conditioned on the known ground truth $a_i^*$. Alternatively, one could first generate reasoning along with predictions, filter out examples that do not match the correct artwork, and repeat this reasoning generation process to obtain a high-quality reasoning dataset~\citep{zelikman2022starbootstrappingreasoningreasoning}. However, due to the difficulty of our prediction task, the pretrained Qwen model achieves low prediction accuracy, which would lead to extensive re-sampling to obtain reasonings that are consistent with the ground-truth and is therefore, difficult to scale. To avoid the expensive re-sampling process, we first provide the model with the true optimal artwork and have it generate an explanation, and have the model predict the best artwork to recommend to the given user conditioned on the model's self-generated justification. We additionally filter out the model's responses that still do not match the ground-truth labels (which accounts for less than 2\% of the model's generations). Despite the model's initial low prediction accuracy, this approach allows us to generate justifications that support the ground truth, which can then be used to construct the reasoning dataset for training. We augment the training data with the generated reasonings, so each tuple $(u_i, x_i, a_i^*)$ is additionally tagged with the corresponding reasoning $r_i$. We then post-train 8B Llama~\citep{weerawardhena2025llama31foundationaisecurityllm8binstructtechnicalreport} using this augmented dataset, with the expected model output starting with ``Reason: ..." followed by the prediction. 

\item \textbf{Direct policy optimization (DPO).} We post-train a 8B Llama model~\citep{weerawardhena2025llama31foundationaisecurityllm8binstructtechnicalreport} using the DPO objective~\cite{rafailov2024directpreferenceoptimizationlanguage}, where the chosen-rejected response pair for each tuple $(x_i, u_i)$ consists of $a_i^*$ (chosen) and a randomly selected alternative $a'_i$ (rejected) from the remaining candidate set. The DPO loss objective\footnote{$\beta$ is a hyperparameter that determines how close the trained model should be to the base model, $\pi_{\text{ref}}$, for maintaining stability.} \begin{equation}
\begin{split}
\mathcal{L}_{\text{DPO}}(\pi_\theta; \pi_{\text{ref}}) 
&= - \mathbb{E}_{i \sim \mathcal{D}} \Bigg[
\log \sigma \Bigg(
\beta \log \frac{\pi_\theta(a_i^* \mid x_i, u_i)}{\pi_{\text{ref}}(a_i^* \mid x_i, u_i)} \\
&\quad - \beta \log \frac{\pi_\theta(a'_i \mid x_i, u_i)}{\pi_{\text{ref}}(a'_i \mid x_i, u_i)}
\Bigg)
\Bigg]
\end{split}
\end{equation}leads the recommendation model to up-weight the likelihood of a preferred response and down-weight the likelihood of a rejected response. This can be more effective than training with only positive examples, as the model explicitly learns to distinguish between positive and negative responses. On the other hand, SFT-only model may learn the instruction-following behavior but does not get to observe how negative examples compare to positive examples despite the structural similarities in their response patterns. DPO can be combined with SFT by first training the base model using the SFT objective and then continuing training with the DPO objective, as suggested by ~\citet{chen2025bootstrappinglanguagemodelsdpo, liu2024understandingreferencepoliciesdirect}. 
\end{itemize}
All models are post-trained using low-rank adaptation (LoRA)~\citep{hu2021loralowrankadaptationlarge}. We conduct a hyperparameter search across different learning rates (e.g., $\{1e-7, 5e-7, 1e-6, 5e-6, 1e-5, 1e-4\}$) and report evaluation results obtained with the best-performing model from the validation set of size 1,000. The evaluation dataset has 5K new user-title pairs.

\section{Experiment \& Results} 
\subsection{How do post-trained LLMs compare to the production model?} Table \ref{tab:main_results} shows that the post-trained models achieve a performance improvement of 3-5\% in terms of IPS compared to the production model. Surprisingly, even the zero-shot model achieves decent performance significantly better than random guessing. This speaks to the power of world knowledge and long-context processing capabilities that the newer language models are increasingly equipped with, which can lead the models to make reasonable predictions about user behavior even though they are not trained on any specific user or title dataset. 

% In table 1, can we briefly explain why we don’t use DPO with reasoning as another comparison method? --> We don't have good reasons for not including this baseline...
\begin{table}
  \caption{We conducted experiments with Llama 8B on a training dataset of 110K examples and evaluated on 5K held-out user-title tuples. The reported values are percentage improvements compared to the production model, which is a non-LLM deep neural network trained on a large-scale production dataset.}
  \label{tab:main_results}
  \resizebox{\columnwidth}{!}{%
  \begin{tabular}{lcc}
    \toprule
    Method & Accuracy  & IPS \\
    \midrule
    Random guess & -74.96\% & -4.59\% \\
    % Production model & 25.84 & 1.047\\
    Zero-shot prediction & -4.22\%  & -0.19\% \\
    SFT & -2.55\% & +2.45\%\\
    \textbf{DPO} & \textbf{+0.91\%} & \textbf{+2.82\%} \\ 
    \textbf{SFT with reasoning from Qwen-32B} & \textbf{+1.41\%} & \textbf{+5.21\%}\\
  \bottomrule
\end{tabular}
}
\end{table}

% \begin{table}
%   \caption{We fine-tune Llama-8B-Instruct on a training dataset of 110K examples and report the results using the best-performing model from the hyperparameter search. The test set contains 5,323 samples of held-out user and title combinations.}
%   \label{tab:main_results}
%   \begin{tabular}{lcc}
%     \toprule
%     Method & Accuracy (\%) & IPS \\
%     \midrule
%     Random guess & 7.04 & 1\\
%     Production model & 15.48 & 1.047\\
%     Zero-shot prediction & 14.84  & 1.045\\
%     SFT & 15.09 & 1.073\\
%     DPO & 15.34 & 1.077 \\ 
%     \textbf{SFT with reasoning from Qwen-32B} & \textbf{15.70} & \textbf{1.103}\\
%   \bottomrule
% \end{tabular}
% \end{table}

\subsection{How sensitive is the model to number versus text based prediction?} 
\begin{figure}[h]
  \centering
  \includegraphics[width=\linewidth]{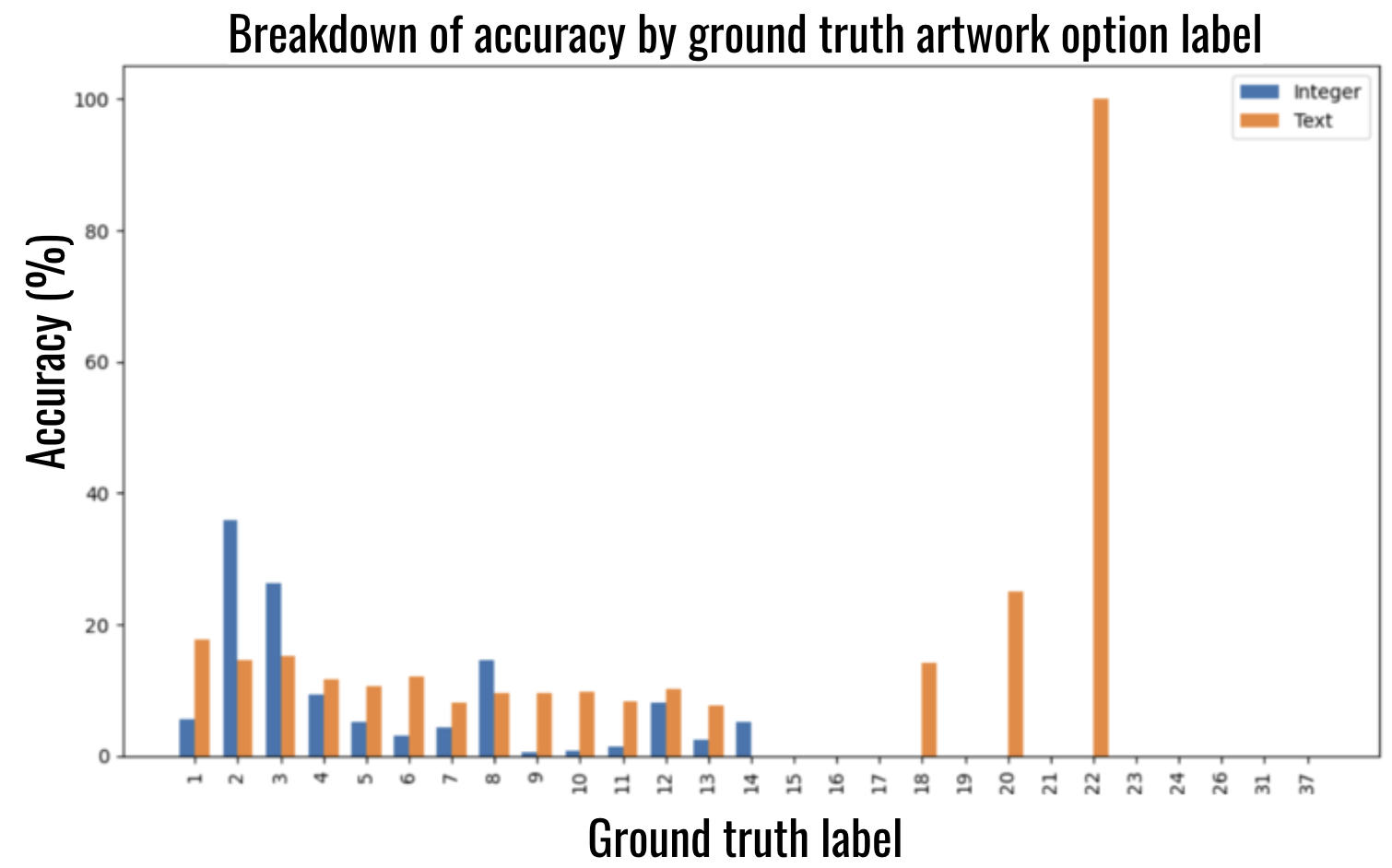}
  \caption{\textbf{A breakdown of accuracy across different ground truth labels to compare the performance of the models outputting the artwork option number versus the artwork caption in full text.} The x-axis shows the ground truth label (artwork option number) and the y-axis shows the model's prediction accuracy for samples with a particular ground truth label. Although the average performance across all examples is higher for the number prediction (blue) compared to the text caption prediction (orange), the breakdown by the ground truth label suggests that number prediction performs poorly for cases with higher numbers for the selected artwork.}\label{fig:breakdown}
  % \Description{A breakdown of accuracy across different ground truth labels to compare the performance of the models outputting the artwork option number versus the artwork caption in full text.}
\end{figure}

Many prior works have observed the model's performance sensitivity to different prompt templates and rephrasings~\citep{berglund2024reversalcursellmstrained, 10.1162/tacl_a_00681, zhuo2024prosaassessingunderstandingprompt} . One particular dimension of sensitivity we investigate is the output formatting: whether it is formatted as an integer (artwork option number) or as a full-text caption describing the artwork. In order to decide on the output format, we conduct an ablation with zero-shot 3B and 8B Llama models comparing their performance on text output versus artwork option number output. Surprisingly, we observe that with the 3B models, the number output outperforms the text output in terms of the evaluation metrics, which is different from the performance of the 8B models (Table \ref{tab:ablation1}). However, a breakdown analysis of the model's accuracy across different ground truth labels (Fig. \ref{fig:breakdown}) reveals the model's tendency to output smaller numbers. In particular, the 3B model prompted to predict the artwork option number shows 0\% accuracy for examples with ground truth labels larger than 15, suggesting that the model's high accuracy is compounded by its bias toward predicting smaller numbers. 

\begin{table}
  \caption{We compare the zero-shot performance of the Llama 3 and 8B models on a test set of 5K user–title pairs to the production model’s performance in terms of percentage difference. The smaller the difference, the closer the zero-shot model’s performance matches the production level.}
  \label{tab:ablation1}
  \resizebox{\columnwidth}{!}{%
  \begin{tabular}{llcc}
    \toprule
   Model size & Output format & Accuracy  & IPS \\
   % Production model & 15.84 & 1.047\\
    \midrule
   3B & Number  & -11.55\% & -6.81\% \\
   3B & Text & -12.76\% & -8.15\%  \\ 
   8B & Number & -7.06\% & -1.73\% \\
   \textbf{8B} & \textbf{Text} & \textbf{-6.52\%} & \textbf{-0.19\%} \\  \bottomrule
\end{tabular}
}
\end{table}

% \begin{table}
%   \caption{We compare the zero-shot performance of the Llama 3 and 8B models on a test set of 5K user–title pairs to the production model’s performance in terms of percentage difference.}
%   \label{tab:ablation1}
%   \begin{tabular}{llcc}
%     \toprule
%    Model size & Output format & Accuracy (\%) & IPS \\
%    % Production model & 15.84 & 1.047\\
%     \midrule
%    3B & Number  & 14.11 & 0.978 \\
%    3B & Text & 13.94 & 0.965  \\ 
%    8B & Number & 14.76 & 1.029 \\
%    \textbf{8B} & \textbf{Text} & \textbf{14.84} & \textbf{1.045} \\  \bottomrule
% \end{tabular}
% \end{table}

\subsection{How does post-training impact different base models?} 
In the previous section, we showed that SFT with reasoning, where the models are trained to output the chosen artwork in text, achieves the highest post-trained performance. Ablations in Table \ref{tab:ablation_change} further test whether this trend holds for different model sizes and training dataset sizes. Specifically, these ablations show that training with a reasoning dataset and outputting text predictions achieves the highest performance gains for both 3B and 8B base models, and for both 10K- and 110K-sized training datasets. As expected, the performance gains from the larger dataset are greater than those from the smaller dataset for the same model size. Surprisingly, we observe that the smaller models obtain larger performance improvements from post-training. However, the final performance of the 8B models still dominates that of the 3B models when compared to the production model's performance. 
% We are interested in empirically answering whether the final performance is constrained by the initial model's quality, i.e., if we have $M$ different models to train with an additional $N$ data points, should we select the best model based on its pre-trained performance, or can a weaker initial model eventually surpass other models after training with a moderately sized (n = 10K) dataset?

\begin{table}
\centering
  \caption{We conducted experiments with Llama 3 \& 8B using LoRA-SFT and evaluated with 5K data points. We report the percentage improvements after post-training compared to the base model of the same type.}
  \label{tab:ablation_change}
  \resizebox{\columnwidth}{!}{
  \begin{tabular}{lllcc}
    \toprule
   Model & Output & Training size & Accuracy & IPS \\
    \midrule
    3B & Number & 10K &  6.71\% & 5.76\% \\
    3B & Text & 10K & 8.38\% & 8.06\% \\
    \textbf{3B} & \textbf{Text + Reason} & \textbf{10K} & \textbf{10.53\%} & \textbf{10.60\%} \\
    8B & Number & 10K &  -0.40\% & 0.67\% \\
    8B & Text & 10K & 1.14\% & 1.52\% \\
    \textbf{8B} & \textbf{Text + Reason} & \textbf{10K} & \textbf{3.05\%} & \textbf{3.02\%} \\
    8B & Number & 110K & 1.64\% & -0.96\% \\
    8B & Text & 110K  & 1.67\% & 3.02\% \\
    \textbf{8B} & \textbf{Text + Reason} & \textbf{110K}  & \textbf{5.63\%} & \textbf{5.40\%} \\
  \bottomrule
\end{tabular}
}
\end{table}

Table \ref{tab:dpo_checkpoint} compares the DPO-trained performance of different checkpoints. While we observe that DPO from the SFT model checkpoint outperforms DPO from the off-the-shelf pretrained model, it does not outperform SFT with reasoning alone. We suspect that the negative examples used to construct the preference pairs may not be clearly distinguished from their positive counterparts. For example, the current training dataset has a single ground-truth optimal artwork which the user has engaged with and liked, while the remaining artworks for the same title are considered negative. However, it is unclear whether a negative example was skipped by the randomized artwork selection algorithm in production, or whether the user (if this artwork were presented to them) would not watch the title due to their mis-specified artwork preference. 

\begin{table}
\caption{We compared the performance of DPO from various checkpoints. All the experiments were conducted with Llama 8B. The reported values are percentage improvements compared to the production model. (Some of the metrics experience performance drops relative to the production model. The smaller the value, the closer the post-trained performance matches the production level.)}
\label{tab:dpo_checkpoint}
\centering
\resizebox{\columnwidth}{!}{
      \begin{tabular}{llcc}
        \toprule
       Training size & SFT checkpoint? & Accuracy & IPS \\
        \midrule
        % Production model & 15.84 & 1.047\\
       10K & \faCheck (SFT without reasoning)   & -4.26\% & +1.42\% \\
       10K & \faCheck (SFT with reasoning)  & -3.73\% & 0\% \\
       110K & \faTimes   & -5.38\% & -0.19\% \\ 
       \textbf{110K} & \faCheck \textbf{(SFT with reasoning)} & \textbf{-3.21\%} & \textbf{+2.82\%}   \\ 
       \bottomrule
    \end{tabular}
    }
\end{table}

%  \begin{tabular}{llcc}
%     \toprule
%    Training size & SFT checkpoint? & Accuracy & IPS \\
%     \midrule
%    10K & \faCheck   & 15.18 & 1.062 \\
%    10K & \faCheck (SFT with reasoning)  & 15.26 & 1.047 \\
%    110K & \faTimes   & 15.01 & 1.045 \\ 
%    \textbf{110K} & \faCheck \textbf{(SFT with reasoning)} & \textbf{15.34} & \textbf{1.077}   \\ 
%    \bottomrule
% \end{tabular}

% \begin{table}
%   \caption{We fine-tune Llama-8B-Instruct or different SFT-trained checkpoints from this base model using the DPO objective on a training dataset of 113,357 examples and report the results using the best-performing model from the hyperparameter search. The test set contains 5,323 samples of held-out user and title combinations.}
%   \label{tab:ablation_dpo}
%   \begin{tabular}{ccl}
%     \toprule
%     Base model & IPS (pre-DPO) & IPS (post-DPO) \\
%     \midrule
%     Untrained & 14.84 \rightarrow 15.01 & 1.045 \rightarrow 1.045 \\
%     SFT-v1 &  & \\
%     SFT-v2 &   & \\
%   \bottomrule
% \end{tabular}
% \end{table}

\section{Discussion \& Limitation} 
In summary, our work introduces the novel recommendation task of predicting personalized artwork for different user-title pairs. Specifically, we provide LLMs with a verbalized representation of a user's interaction history, which reveals their content preferences, and ask the model to predict which artwork caption from a candidate set is most likely to appeal to the given user. We leverage the LLM's natural language reasoning capabilities and show that the model's world knowledge augmented with post-training on 110K user data achieves a 5\% performance improvement compared to the production model. Additionally, in our hyperparameter search, we observe that smaller learning rates (e.g., 1e-7 $\sim$ 1e-6) generally help, which is consistent with findings in ~\citet{pareja2024unveilingsecretrecipeguide}. 

Our work points to several promising directions for personalized recommendation:

\textbf{Improving models with reasoning.} Augmenting the recommendation model's reasoning capabilities is another promising direction. Based on the improvements we observed with SFT with reasoning compared to standard SFT, we expect DPO with reasoning to further improve the model's performance and also engender users' trust in AI systems by providing interpretable explanations for the model's personalized recommendations.

\textbf{Leveraging multimodal LLMs.} While our work represents artworks with captions, recent advances in multimodal LLMs (e.g., GPT-5~\citep{gpt5}, Gemini 2.5~\citep{gemini2.5}) suggest the possibility of directly embedding images into the prompt, thereby avoiding information loss from the transformation.

\textbf{Personalizing various aspects of content recommendation.} Finally, we encourage future work to consider extending this framework to other components of content personalization beyond artworks (e.g., synopses, trailers), or even to directly recommend new artworks or novel features of an artwork in natural language, which service providers can then use for creative artwork generation tailored to individual users' tastes and interests.

\bibliography{example_paper}
\bibliographystyle{icml2025}
\clearpage

\end{document}